\documentclass[doublespacing]{elsart}
\usepackage{epsfig}

\usepackage{amssymb}

\begin{document}

\begin{frontmatter}

\title{Micromechanical electrometry of single-electron transistor 
island charge}

\author{M. P. Blencowe,\corauthref{cor}}
\author{Yong Zhang}
\address{Department of Physics and Astronomy, Dartmouth College,
Hanover, New Hampshire 03755}
\corauth[cor]{Corresponding author:  Fax: +(603)-646-1446,
email: m.blencowe@dartmouth.edu}

\begin{abstract}
    We consider the possibility of using a micromechanical gate 
    electrode located just above the island of a single-electron 
    transistor to measure directly the fluctuating  island charge due to 
    tunnelling electrons. 
\end{abstract}

\end{frontmatter}

Single-electron transistors (SETs) \cite{fulton} have proven to be 
exquisitely sensitive electrometers, with the best reported lower noise values 
of about $10^{-5}~e/\sqrt{\rm Hz}$ \cite{starmark,krupenin}. Further 
improvements in sensitivity require an understanding of the origins of 
the background charge fluctuations (see, e.g., Ref.~\cite{starmark}, 
and references therein) which dominate over the intrinsic tunnelling shot noise 
of current devices. By contrast, there has been little work on 
probing directly the intrinsic noise of the SET itself.
In a relevant investigation \cite{heij}, a superconducting SET 
was coupled capacitively to a SET island in order to measure the 
charge state of the island.   
Such investigations provide additional valuable tests of our 
understanding of the 
SET dynamics, as represented by the orthodox theory \cite{averin}.  

In the present work we consider the possibility of using 
micromechanical electrometry to measure directly the 
fluctuating SET island charge due to 
electrons tunnelling onto and off the island \cite{cleland}. 
A cantilever gate electrode located directly above the SET island 
with displacement measured using either the magnetomotive method 
\cite{mohanty} or fibre optic interferometer \cite{mamin} might serve 
as the electrometer.     

We first derive an order of magnitude estimate of the spectral density of the 
force acting on the 
cantilever due to the fluctuating island charge. 
In terms of the SET source-drain voltage $V_{\rm{sd}}$, the gate voltage 
$V_{g}$ and the number $n$ of excess electrons on the SET island, the 
electrostatic energy stored in the gate capacitor $C_{g}$ is  
\[
U=\frac{C_{g}}{2 C_{\Sigma}^{2}}\left[C(V_{\rm{sd}} -2V_{g}) -ne\right]^{2},
\]
where we assume a symmetric SET with tunnel junction capacitances 
$C_{1}=C_{2}=C$, effective resistances  
$R_{1}=R_{2}=R$, total capacitance $C_{\Sigma}=2C+C_{g}$ and we 
have neglected a possible background island  charge since it will not 
affect the final result. For small displacements of the cantilever 
gate electrode, $C_{g}\approx C_{g}^{0} (1-x/d)$, where $d$ is the 
cantilever electrode-SET island gap, and the force on the cantilever is
\[F=-\frac{\partial U}{\partial 
x}=\frac{1}{d}\left(\frac{2C-C_{g}}{2C+C_{g}}\right) 
U,\]
where we have dropped the superscript `$0$' on $C_{g}$.
At a given source-drain current peak 
maximum,  $n$, $V_{g}$ and $V_{\rm{sd}}$ satisfy the condition
\[e (n+1/2)= C_{g} (V_{g}-V_{\rm{sd}}/2)
\] 
and the probabilities at any given time that there are either $n$ or 
$n+1$ electrons on the island are approximately equal to $1/2$, 
provided the current peaks are 
well-separated in gate voltage. The force noise is also a maximum at 
a current peak maximum; we will restrict ourselves to evaluating this 
force noise maximum which takes a particularly 
simple form. An estimate for the force noise is 
$S_{F}=[F(n+1)-F(n)]^{2}/(I_{\rm{sd}}/e)$, where $I_{\rm{sd}}= 
V_{\rm{sd}}/4R$ is the peak maximum source-drain current. Substituting 
$V_{g}$ and $V_{\rm{sd}}$ for $n$ using the above condition we 
obtain
\[S_{F}=\frac{4 e^{3} R}{V_{\rm{sd}}}
\left[\frac{C_{g}(2C-C_{g}) (V_{g}-V_{\rm{sd}}/2)}
{d C_{\Sigma}^{2}}\right]^{2}.
\]
A full derivation of $S_{F}(\omega)$ using the orthodox theory along 
the lines of Ref.~\cite{korotkov} yields the same result at 
$\omega=0$ and at a current peak maximum, differing only by an overall factor of 
$1/4$~\cite{zhang}.

The minimum detectable force is set by the intrinsic thermomechanical 
noise of the cantilever. For a cantilever with geometry $l\times 
w\times t\ ({\rm length}\times{\rm width}\times{\rm thickness})$, 
quality factor $Q$, 
Young's modulus $E$ and mass density $\rho$, the thermomechanical 
force noise is 
\[ S_{F}^{{\rm th}~1/2}=\left(\frac{w t^{2}}{l Q}\right)^{1/2} 
(E\rho)^{1/4}(k_{B}T)^{1/2}.
\]
Provided there are no constraints on frequency, sensitive force 
detection therefore requires long and thin cantilevers with large quality 
factors~\cite{rugar}. The current best sensitivity is about $0.8~{\rm 
aN}$ in a $1~{\rm Hz}$ bandwidth~\cite{mamin}.   
For the readily achievable values $R=50~{\rm k}\Omega$, $C=0.25~{\rm fF}$, 
$C_{g}=0.1~{\rm fF}$ (corresponding to a $1~\mu{\rm m}^{2}$ plate area 
and $0.1~\mu{\rm m}$ plate gap $d$) and $V_{\rm{sd}}=0.1~{\rm mV} (=0.4 
e/C_{\Sigma})$,  the SET force noise is
\[S_{F}^{1/2}\approx 1.5 V_{g}~{\rm aN}/\sqrt{\rm{Hz}}
\]
for $V_{g}\gg V_{\rm{sd}}$. Thus the SET force noise would be 
detectable for $V_{g}\sim 1~{\rm V}$. However, for such 
voltages, pull-in of the cantilever to the substrate surface would likely occur.
A possible way to avoid this problem is to orient the cantilever 
perpendicular to the surface \cite{rugar}.

Another possible approach is to user a shorter, stiffer cantilever 
and apply a gate voltage $V_{g}=V_{g}^{0}+V_{\rm ac} \cos (2\pi\nu 
t)$, with $V_{g}^{0}\sim 1~{\rm V}$ chosen such that $I_{\rm{sd}}$ is 
at a given peak maximum, $V_{\rm ac}=e/2 C_{g}=0.8~{\rm mV}$, and the 
frequency $\nu$ chosen to be smaller than the cantilever damping rate.
By measuring both the amplitude and phase of the cantilever 
displacement and averaging over sufficiently many cycles \cite{mohanty}, the 
dependence of the SET force noise on gate voltage for a given current 
peak can be determined.

Further work needs to be done to determine the response of the 
cantilever electrode to the fluctuating SET island charge. In 
particular, the backaction of the cantilever on the SET must also be 
taken into account which requires solving for the dynamics of coupled 
SET-cantilever system. This will be the subject of a future work.   

We thank Keith Schwab for discussions which led to the present work. 
This work was supported in part  by the NSA
and ARDA under ARO contract number DAAG190110696, and by an award from 
Research Corporation.

\end{document}